\documentclass[sigconf]{acmart}
\usepackage{listings} 
\usepackage{graphicx}
\usepackage{subfigure}
\usepackage{booktabs}
\usepackage{todonotes}
\usepackage{amsmath,amsfonts,amsthm}
\usepackage{textcomp}
\usepackage{color,xcolor,colortbl}
\usepackage{multirow}
\usepackage{enumitem}
\usepackage{url}
\usepackage[ruled, linesnumbered]{algorithm2e}
\usepackage{bm}
\usepackage{diagbox}

\usepackage{soul} 

\AtBeginDocument{%
  \providecommand\BibTeX{{%
    \normalfont B\kern-0.5em{\scshape i\kern-0.25em b}\kern-0.8em\TeX}}}

\def\numx#1e#2{{#1}\mathrm{e}{#2}}

\copyrightyear{2023}
\acmYear{2023}
\setcopyright{rightsretained}
\acmConference[WWW '23]{Proceedings of the ACM Web Conference 2023}{April 30-May 4, 2023}{Austin, TX, USA}
\acmBooktitle{Proceedings of the ACM Web Conference 2023 (WWW '23), April 30-May 4, 2023, Austin, TX, USA}
\acmDOI{10.1145/3543507.3583345}
\acmISBN{978-1-4503-9416-1/23/04}

\begin{document}

\definecolor{Gray}{gray}{0.85}

\title{BERT4ETH: A Pre-trained Transformer for Ethereum Fraud Detection}

\author{Sihao Hu}
\affiliation{
  \institution{National University of Singapore}
  \institution{Georgia Institute of Technology}
  \country{}
}
\email{sihaohu@gatech.edu}

\author{Zhen Zhang}
\affiliation{%
  \institution{National University of Singapore}
  \country{}
}
\email{zhen@nus.edu.sg}

\author{Bingqiao Luo}
\affiliation{%
  \institution{National University of Singapore}
  \country{}
}
\email{luo.bingqiao@u.nus.edu}

\author{Shengliang Lu}
\affiliation{%
  \institution{National University of Singapore}
  \country{}
}
\email{lusl@nus.edu.sg}

\author{Bingsheng He}
\affiliation{%
  \institution{National University of Singapore}
  \country{}
}
\email{hebs@comp.nus.edu.sg}

\author{Ling Liu}
\affiliation{%
  \institution{Georgia Institute of Technology}
  \country{}
}
\email{ling.liu@cc.gatech.edu}

\begin{abstract}

As various forms of fraud proliferate on Ethereum, it is imperative to safeguard against these malicious activities to protect susceptible users from being victimized.
While current studies solely rely on graph-based fraud detection approaches, it is argued that they may not be well-suited for dealing with highly repetitive, skew-distributed and heterogeneous Ethereum transactions.
To address these challenges, we propose BERT4ETH, a universal pre-trained Transformer encoder that serves as an account representation extractor for detecting various fraud behaviors on Ethereum. 
BERT4ETH features the superior modeling capability of Transformer to capture the dynamic sequential patterns inherent in Ethereum transactions, and addresses the challenges of pre-training a BERT model for Ethereum with three practical and effective strategies, namely repetitiveness reduction, skew alleviation and heterogeneity modeling. 
Our empirical evaluation demonstrates that BERT4ETH outperforms state-of-the-art methods with significant enhancements in terms of the phishing account detection and de-anonymization tasks. The code for BERT4ETH is available at: \url{https://github.com/git-disl/BERT4ETH}.

\end{abstract}

\maketitle

\section{Introduction}

As a decentralized computing platform, Ethereum empowers its developers to create a variety of decentralized applications (DApps). Despite the substantial engagement garnered within the cryptocurrency sphere, Ethereum has also become a hub for a wide range of fraudulent activities, such as phishing scams~\cite{trans2vec}, pump-and-dump schemes~\cite{hu2022sequence}, Ponzi schemes~\cite{chen2018detecting}, ICO scams~\cite{bian2018icorating}, money laundering~\cite{victor2021detecting}, and bot arbitrage~\cite{flashboy}, \textit{etc}.

Many recent studies~\cite{trans2vec,watching,lin2020modeling,lin2020t,chen2020phishing,li2022ttagn} employ graph representation learning techniques for fraud detection on Ethereum. Although it is intuitive to represent the interactions between accounts as a graph, it is argued that they have the following limitations: 

(\romannumeral1) Graph is not appropriate for capturing the sequential pattern inherent in transactions. Ethereum transactions are high repetitive, indicating the presence of multi-edges between nodes. Current graph-learning methods~\cite{trans2vec,deepwalk,diff2vec} integrate multi-edges to a single edge to facilitate graph computations. However, the discarded sequential information is essential for characterizing user behaviors for tasks such as de-anonymization. 

(\romannumeral2) Graph Neural Networks (GNNs), especially on the highly skew-distributed Ethereum data~\cite{lee2020measurements}, can suffer from noise when the number of convolution hops increases~\cite{Turbo}, given that Ethereum accounts are often connected to highly popular accounts. However, limiting the number of convolution hops can restrict the capabilities of GNNs, as the number of hops is typically equivalent to the depth of layers in conventional GNNs~\cite{li2022ttagn,gcn,graphsage,gat}.

(\romannumeral3) Existing studies primarily target on individual fraud detection tasks in an end-to-end training manner. In light of the successes of Transformer pre-training techniques in NLP~\cite{devlin2018bert,transformer}, we believe that a pre-trained Transformer can support various fraud detection tasks with minimal adaptations needed.

To address the limitations discussed above, we introduce a pre-trained model that offers a universal solution for various fraud detection tasks on Ethereum. BERT4ETH features the superior sequential modeling capability of Transformer~\cite{transformer} and the pre-training paradigm of BERT~\cite{devlin2018bert}. In this paper, we first present the architecture of BERT4ETH, with a specific focus on the integration of Transformer into the Ethereum contenxt. BERT4ETH serves as a sequential encoder, capable of extracting representation vectors for user accounts based on their transaction histories. Second, we introduce the Masked Address Prediction (MAP) task, which involves randomly masking addresses (accounts) in transaction sequences and requiring the model to predict the masked addresses. The MAP task forces the model to learn the relationship between addresses (accounts) in transaction sequences.

However, three characteristics of Ethereum pose challenges for pre-training: (\romannumeral1) \textit{Repetitiveness}: High repetitiveness prevents BERT4ETH from learning meaningful representations through the MAP task, because label information is very likely leaked from unmasked addresses to masked addresses. (\romannumeral2) \textit{Skewed distribution}: The frequency of occurrence of addresses follows the power-law distribution~\cite{lee2020measurements}, with a small number of popular addresses proliferating in the majority of transaction sequences. This reduces the distinctiveness of representations that is what fraud detection covet most. (\romannumeral3) \textit{Heterogeneity}: Ethereum transactions include various types of interactions (Ether/token transfer, contract call) between different types of accounts, creating a challenge in modeling the heterogeneity and uncovering meaningful patterns behind transactions.

We tackle the above challenges by equipping BERT4ETH with the following three strategies:
\begin{itemize}[leftmargin=*]
\item \textit{Repetitiveness reduction:} To counter the label leakage problem in pre-training, we first aggregate continuously repetitive transactions while preserving the sequential order. Second, we propose two alternative effective strategies: adopting a \textit{high} masking ratio (80\%) or a \textit{high} drop out ratio (80\%) during pre-training. These tactics create a task that cannot be easily extrapolated by the high repetitiveness.

\item \textit{Skew alleviation:} We emphasize the distinctiveness by sampling high-frequency addresses as negative samples in a contrastive loss function adopted for pre-training. Optimizing the contrastive loss equals to alleviating the negative impact of high-frequency addresses. Additionally, we propose an intra-batch sharing strategy for negative samples, which allows an extremely high negative-to-positive ratio to alleviate the skewness, and decreases the overlap of negative sets because of frequency-aware sampling.

\item \textit{Modeling heterogeneity:} BERT4ETH captures fine-grained transaction information by the embedding technique and models sequential pattern by a Transformer encoder. Furthermore, (\romannumeral1) we separate transaction sequences into in/out-type sub-sequences and model them individually because fraudsters can be more easily distinguished in specific types of in- or out-transactions; (\romannumeral2) We propose a log encoder to integrate the trace of ERC-20 token transfer without introducing noise, since token transfer trace cannot be captured by normal transactions.

\end{itemize}

Extensive experiments conducted on two crucial fraud detection tasks show that BERT4ETH significantly advances the state-of-the-art performance, achieving a $F_1$ improvement of \textbf{21.61} absolute percentage (AP) for phishing detection and Hit Ratio@1 improvement of \textbf{13.54} and \textbf{21.57} AP for de-anonymization on the ENS and Tornado (0.1ETH) datasets, respectively.

\noindent \textbf{Contributions:} To summarize, the contributions are as follows:
\begin{itemize}[leftmargin=8 pt]

\item We present BERT4ETH, a pre-trained Transformer that provides a universal solution for various Ethereum fraud detection tasks.

\item We equip BERT4ETH with three effective strategies to generate robust and expressive representations, given repetitive, skew-distributed and heterogeneous Ethereum transaction.

\item BERT4ETH significantly advances the state-of-the-arts on two important fraud detection tasks. As a side contribution, we make available the code and dataset.

\end{itemize}

\section{Related Work and Background}

\subsection{Ethereum Representation Learning}

Previous studies have primarily focused on graph-based methods for Ethereum account representation learning, which can be classified as DeepWalk-based and GNN-based methods. 

\textit{DeepWalk-based} method: Trans2Vec~\cite{trans2vec} is proposed for the phishing account detection task, which integrates temporal and amount information of transactions into its random walk process, making the proximity of learned node representations reflects the relationship between accounts. Other works, such as~\cite{lin2020modeling,lin2020t}, also take inspiration from Trans2Vec. For the task of de-anonymization, which aims to identify two accounts belonging to a single user based on the proximity of account representations, Beres et al.\cite{watching} evaluate 11 graph learning methods on ground-truth pairs collected from the ENS and Tornado coin-mixers. Among them, Diff2Vec\cite{diff2vec} and Role2Vec~\cite{role2vec} are considered the state-of-the-art methods.

\textit{GNN-based} method: Shen et al.~\cite{shen2021identity} utilize Graph Convolution Network (GCN)~\cite{gcn} to classify accounts into "normal," "phisher," and "bot" categories based on inferred identities. Zhou et al.~\cite{zhou2022behavior} propose HGATE, a hierarchical graph attention encoder that integrates features from both node-level and subgraph-level to enhance phishing detection performance. Li et al.~\cite{li2022ttagn} propose TTAGNN, a GNN that fuses multiple temporal edges by using a LSTM network, and learns node embeddings through a Graph Attention Network (GAT)~\cite{gat}. A graph auto-encoder is employed to generate a self-supervised signal for representation learning, with a LightGBM model adopted for the phishing account detection task.

\subsection{Transformer \& BERT}
Transformer~\cite{transformer} is a sequence-to-sequence machine translation model that introduce the groundbreaking self-attention mechanism for capturing the relationship between word tokens. BERT~\cite{devlin2018bert} proposes Masked Language Modeling (MLM) and Next Sentence Prediction (NSP) task to pre-train the Transformer encoder in a bidirectional context. It advances the state-of-the-art on eleven NLP tasks with a significant enhancement, and has inspired a number of variants such as ALBERT~\cite{lan2019albert}, RoBERTa~\cite{liu2019roberta} and XLNet~\cite{yang2019xlnet}. Moreover, the pre-training paradigm of BERT makes it can be easily extended to various downstream tasks. In light of the success of Transformer and BERT pre-training, our research aims to take advantages of their superior capabilities for Ethereum fraud detection.

\subsection{Terminology}

\label{sec:term}
\textit{Externally Owned Account (EOA):} EOAs are accounts controlled by users who own their private keys, allowing them to initiate external transactions for transferring cryptocurrency or triggerring smart contracts. This study focuses on modeling the transactions initiated by EOAs as they are under human control.

\textit{Contract Account:} Contract accounts are self-executing computer programs deployed on the Ethereum network. Ethereum allows encoding of arbitrary contract functionality. While contract accounts cannot issue external transactions, they can initiate internal transactions.

\textit{External Transaction:} An external transactions initiated exclusively by an EOA, either transfers cryptocurrency to other accounts or call a contract account to trigger its execuation. In comparison, there are internal transactions initiated by smart contracts to execute complex logic. In this study, the term "transaction" specifically refers to external transactions, unless specified otherwise.

\textit{Token:} Tokens are digital assets that can be programmed to serve various functions, such as functioning as as a currency, granting access, voting, providing identity and utility. Currently, the majority of tokens are built upon the ERC-20 standard~\cite{chen2020traveling}.

\vspace{0.1cm}
\section{Motivation}
\label{sec:motivation}

We introduce three challenges/characteristics of Ethereum that motivate us to design a new BERT-based model.

\textit{Repetitiveness:} Ethereum transactions are highly repetitive. Statistics indicate that there are 48.4\% of transactions share the same receiver with its one prior transaction initiated by the same sender, suggesting that Ethereum users exhibit a tendency to repeatedly interact with the same accounts. However, the pre-training task of BERT4ETH is vulnerable to high repetitiveness: label information can leak from unmasked tokens to masked but repetitive ones, hindering BERT4ETH to capture meaningful co-occurrences between addresses, a phenomenon we refer to as the \textit{label leakage} problem.

\begin{figure}[tbp]
\vspace{-0.2cm}
\includegraphics[width=5.5cm]{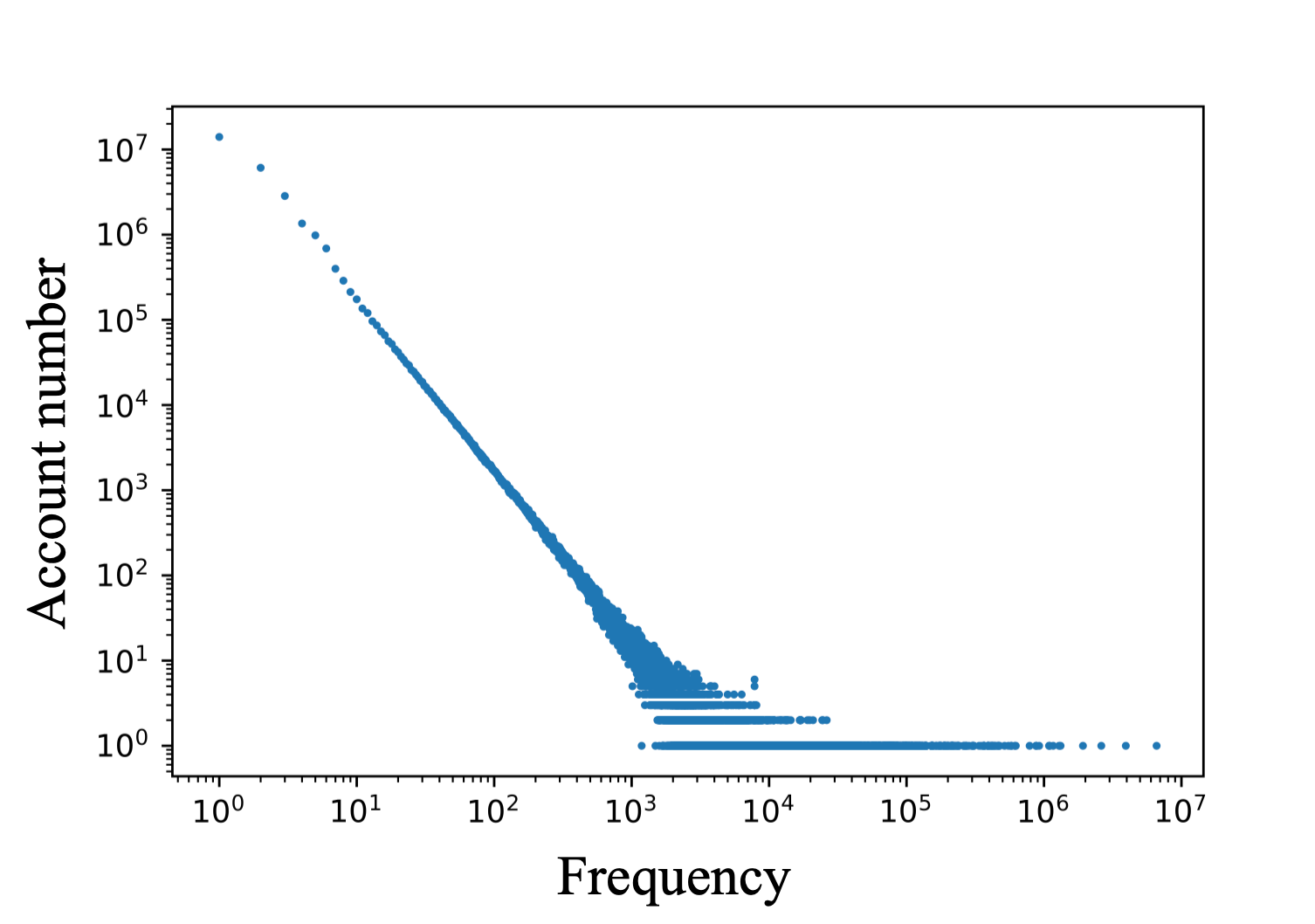}
\caption{The frequency of occurrence of account (address) follows a power-law distribution.}
\vspace{0.1cm}
\label{fig:power_law}
\end{figure}

\textit{Skew-distributed:} As shown in Figure~\ref{fig:power_law}, the frequency of occurrence of accounts (addresses) follows the power-law distribution~\cite{lee2020measurements,zhao2021temporal}, which means a small number of high-frequency accounts proliferate in the majority of transactions. This characteristic presents the difficulty for representation learning, as it can diminish the distinctiveness of representations: two accounts that interacted with popular accounts like Uniswap are likely be closely located in the latent space, even if they are completely irrelevant.

\textit{Heterogeneous:} In Ethereum, there exist different types of accounts, transactions and functionalities associated with calling contract accounts. As compared to human languages, the heterogeneity present in transactions makes it more challenging to discern meaningful patterns and determine the most important elements of information. Given that various downstream fraud detection tasks depend on different aspects of information, we aims to preserve heterogeneity as much as possible during the pre-training phase.

\section{BERT4ETH}

In this section, we present the design of BERT4ETH, along with three strategies aimed at addressing the above-mentioned challenges, dubbed Repetitiveness Reduction (RR), Skew Alleviation (SA) and Modeling Heterogeneity (MH).

\subsection{Transaction Sequence}

\subsubsection{\textbf{Data Collection}}

We deployed an Ethereum node by Geth and utilized Ethereum-ETL to extract structured tabular data from archived raw data. The table schema used in this paper is at \url{https://ethereum-etl.readthedocs.io/en/latest/schema}, where \textit{transaction.csv} is the external transaction file and \textit{trace.csv} is the log file.

\subsubsection{\textbf{Sequence Generation}} For an EOA with address $A_{0}$, we collect all the transactions which it was either the initiator or the receiver, and sort transactions in \textit{descending} order based on their timestamps. For each transaction, we collect four features, \textit{i.e.}, address, timestamp and amount, account type, and in/out type. 
In/out type feature indicates whether the transaction is received or initiated by $A_0$, account type indicates whether the account is EOA or contract, amount is the value of transferred amount, and timestamp denotes the transaction time. 
Subsequently, we insert a dummy self-transaction at the head of the sequence. The address feature of the self-transaction is set to $A_{0}$ (self-address), and all the other features are set to "Null", to differentiate it from normal transactions.

\textbf{Transaction De-duplication (RR\#1):} The first strategy of repetitiveness reduction is transaction de-duplication, aiming to reduce continuous repetitiveness. Continuous repetitiveness refers to transactions that interacts with the same addresses continuously in a sequence. First, we eliminate failed transactions as user may initiate several failed transactions before a final one successfully executed. Second, we aggregate continuous repetitive transactions that have same address, same in/out type and initiated within 72 hours into one, by summing up their transaction amounts and tracking the number of transactions. The timestamp of the aggregated transaction is set to the first timestamp of the original transactions. By adopting de-duplication strategy, we lower the repetitiveness ratio from 48.0\% to 14.3\%, while still preserving the order of the original sequence.

\subsection{Model Architecture}

\subsubsection{\textbf{Embedding Layer}}

\begin{figure}[tbp]
\includegraphics[width=7.0cm]{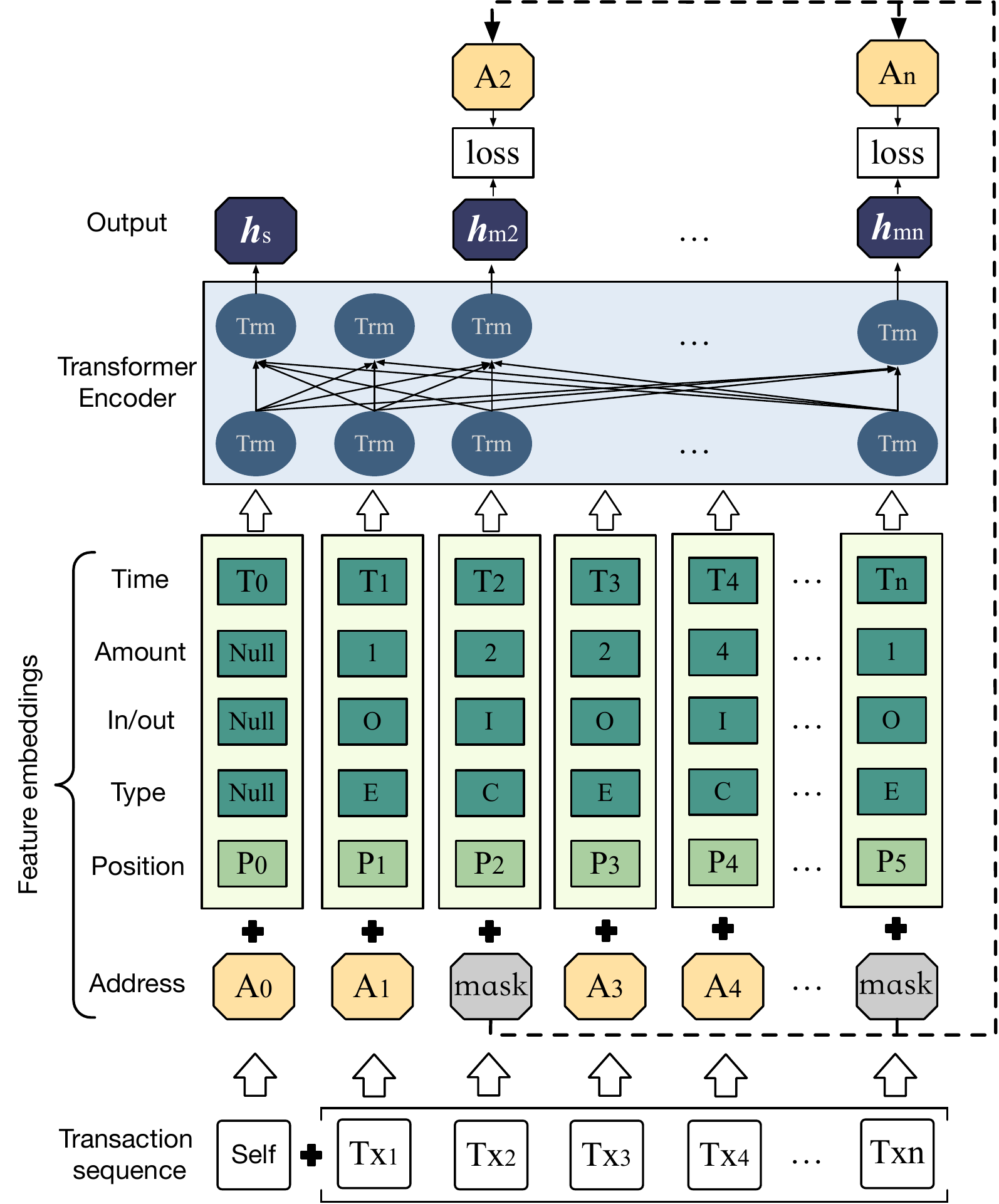}
\vspace{0.25cm}
\caption{The framework of BERT4ETH pre-training. After a transaction sequence is generated, we select a portion of transactions to replace their addresses with [MASK] and feed the sequence to the model to predict masked addresses.}
\vspace{0.1cm}
\label{fig:bert_model}
\end{figure}

As illustrated in in Figure~\ref{fig:bert_model}, seven features are generated for each transaction, including address, account type, in/out type, amount, count, timestamp and position. Since amount and count are no-categorical features, we use binning to categorize them. The position index is ranked from 0 to $N-1$.

First, we adopt the embedding technique to encode features to make the model aware of the transaction information. Specifically, for the $i$-th transaction in the sequence, its transactions features are passed through embedding layers to generate the corresponding feature embeddings, which are then summed to obtain its initial transaction representation $\boldsymbol{h}_{i}^{(0)} \in \mathbb{R}^{d}$.
Next, we stack the initial transaction representations to form a matrix $\boldsymbol{H}^{(0)}=[\boldsymbol{h}_{0}^{(0)}, \boldsymbol{h}_{1}^{(0)}, ..., \boldsymbol{h}_{N-1}^{(0)}] \in \mathbb{R}^{N \times d}$ that encompasses all the information of the transaction sequence. 

\subsubsection{\textbf{Transformer Encoder}} For a sequence, BERT4ETH takes $\boldsymbol{H}^{(0)}$ as the input, and passes it through the Transformer encoder consisting of $L$ Transformer layers. Each Transformer layer contains two sub-layers, an Attention sub-layer and a Position-wise Feed-Forward sub-layer. We formalize a Transformer layer as follows:
\begin{equation}
\boldsymbol{H}^{(l)} =\operatorname{Attention}\left(\boldsymbol{H}^{(l)} \boldsymbol{W}_{Q}^{(l)}, \boldsymbol{H}^{(l)} \boldsymbol{W}_{K}^{(l)}, \boldsymbol{H}^{(l)} \boldsymbol{W}_{V}^{(l)}\right)
\label{eq:1}
\end{equation}
\begin{equation}
\operatorname{Attention}(Q, K, V)=\operatorname{softmax}\left(\frac{Q K^{\top}}{\sqrt{d}}\right) V
\label{eq:2}
\end{equation}
\begin{equation}
\boldsymbol{H}^{(l+1)} = [\operatorname{FFN}(\boldsymbol{h}_{1}^{(l)});\cdots; \operatorname{FFN}(\boldsymbol{h}_{t}^{(l)})]
\label{eq:3}
\end{equation}
\begin{equation}
\operatorname{FFN}(\boldsymbol{x}) = \operatorname{GELU}(\boldsymbol{x} \boldsymbol{W}_{1}^{(l)} + \boldsymbol{b}_{1}^{(l)}) \boldsymbol{W}_{2}^{(l)} + \boldsymbol{b}_{2}^{(l)}
\label{eq:4}
\end{equation}
where the projection matrices $\boldsymbol{W}_{Q}^{(l)}$, $\boldsymbol{W}_{K}^{(l)}$, $\boldsymbol{W}_{V}^{(l)}$, $\boldsymbol{W}_{1}^{(l)}$, $\boldsymbol{W}_{2}^{(l)} \in\mathbb{R}^{d\times d}$, and bias vectors $\boldsymbol{b}_{1}^{(l)}$ and $\boldsymbol{b}_{2}^{(l)} \in\mathbb{R}^{d\times 1}$ are trainable parameters for the $l$-th Transformer layer. To facilitate description, we omit the multi-head~\cite{devlin2018bert}, ResNet~\cite{resnet} and batch normalization~\cite{batchnormalization}, but they are adopted in practice. 

After $L$-layer successive calculation, the Transformer encoder produces a matrix $\boldsymbol{H}^{(L)}=[\boldsymbol{h}_{0}^{(L)}, \boldsymbol{h}_{1}^{(L)}, ..., \boldsymbol{h}_{N-1}^{(L)}] \in \mathbb{R}^{N \times d}$, where $\boldsymbol{h}_{i}^{(L)}$ is the final representation of the $i$-th transaction, which encodes not only its own information but also the bi-directional context information.

\subsection{Pre-training}

\subsubsection{\textbf{Masked Address Prediction}} 
MAP is derived from the Masked Language Modeling (MLM) of BERT, which involves a \textit{Cloze} test that requires the model to predict the masked addresses in a transaction sequence, as shown in Figure~\ref{fig:bert_model}. In BERT4ETH, a certain percentage ($k$\%) of transactions within the sequence are selected and their addresses are replaced with the special token [MASK]. The masked sequence is then passed through the embedding and Transformer layers as described before. For a masked transaction, its final transaction representation $\boldsymbol{h}^{(L)}_m$, which encodes its bidirectional contextual information, is used to predict its masked address. 

The original BERT predicts masked word tokens through the calculation of probabilities across all tokens (around 30K in number). However, when applied to Ethereum, it is infeasible to calculate $\rm{softmax(\cdot)}$ across all the addresses as there are up to billions of addresses in Ethereum. Therefore, we adopt a contrastive loss calculated over a positive address and random sampled negative addresses as the objective function for pre-training:
\begin{equation}
    L = -\frac{1}{|\mathcal{M}|} \sum_{m \in \mathcal{M}} \operatorname{log}\left (
    \frac{\operatorname{exp}(\boldsymbol{h}_m^{\mathrm{T}} \cdot \boldsymbol{a}_p)}{\operatorname{exp}(\boldsymbol{h}_m^{\mathrm{T}} \cdot \boldsymbol{a}_p) + \sum_{n \in \mathcal{N}} \operatorname{exp}(\boldsymbol{h}_m^{\mathrm{T}} \cdot \boldsymbol{a}_n)} \right)
\label{eq:loss}
\end{equation}
where $\mathcal{M}$ is the set of masked addresses in a sequence, $\boldsymbol{h}_m$ is $\boldsymbol{h}^{(L)}_m$ that encodes unmasked contextual information. For each sequence, we samples a negative set $\mathcal{N}$. 
$\boldsymbol{a}_p$ is the address embedding of its masked address, which we refer to the positive embedding, and $\boldsymbol{a}_{n}$ is the address embedding of a negative address from the negative set $\mathcal{N}$. Here we reuse the address embedding layer to prevent introducing new parameters.
Optimizing Eq.~\ref{eq:loss} is essentially equivalent to encouraging $\boldsymbol{h}_m$ be closer to $\boldsymbol{a}_p$, and away from $\boldsymbol{a}_n$ in the hidden space.

\begin{figure}[tbp]
\centering
\includegraphics[width=8.0cm]{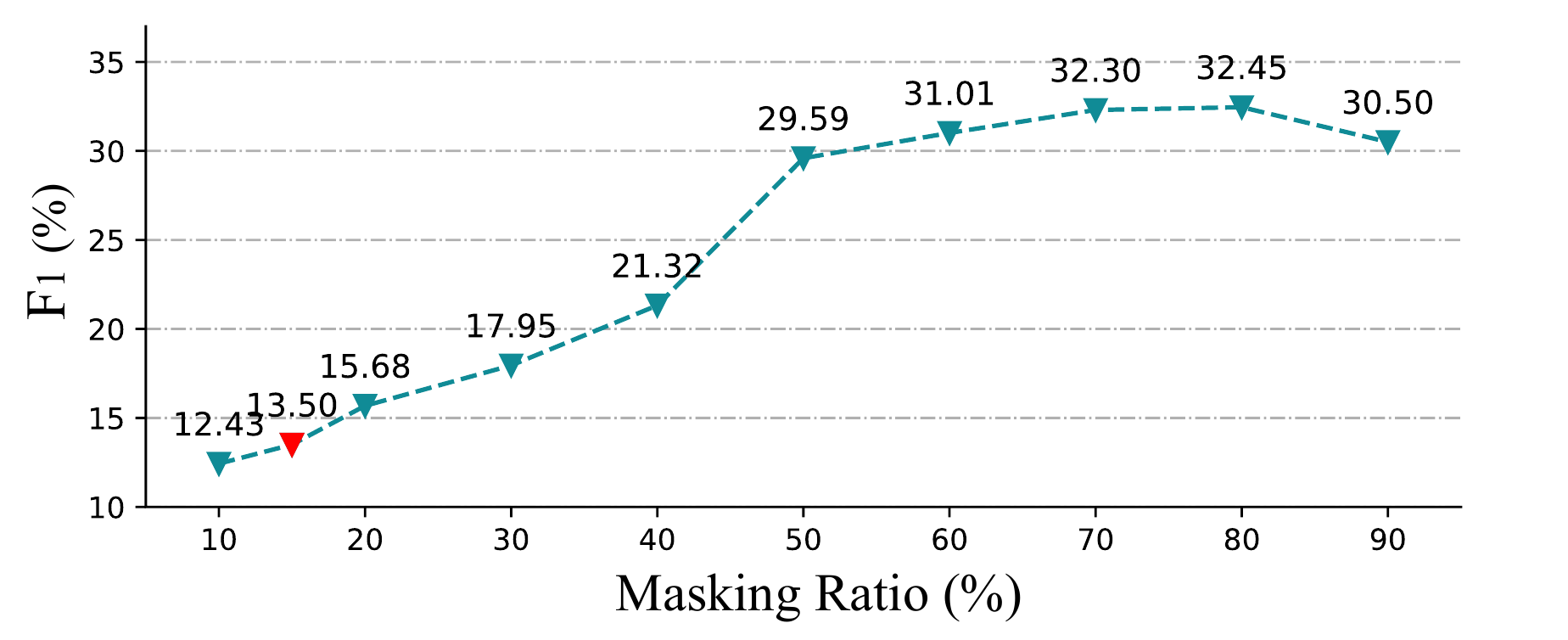}
\vspace{0.1cm}
\caption{Testing $F_1$ of phishing account detection \textit{w.r.t.} different masking ratios.
A high masking ratio (80\%) works significantly better than the original ratio (15\%).}
\vspace{0.1cm}
\label{fig:mask_ratio}
\end{figure} 

\textsc{Multi-hop modeling:} From a graph perspective, the address embedding captures up to two-hop neighborhood information during MAP pre-training ($A_0$ and $A_1$ are one-hop neighbors, while $A_1$ and $A_3$ are two-hop neighbors). After pre-training, the Transformer encoder extracts the account representation of $A_0$ using its transaction sequence as input. This explicitly includes the address embedding of its one-hop neighbors and implicitly captures the information from the two-hop neighbors of $A_0$'s immediate neighbors. In total, BERT4ETH captures three-hop neighborhood information, as will be discussed in our experiment (Section~\ref{sec:multihop_modeling}).

\subsubsection{\textbf{Repetitiveness Reduction}}
\label{sec:redundancy}

The issue of high repetitiveness poses a risk of label leakage for the MAP task, which can negatively impact the effectiveness of pre-training. For instance, if BERT4ETH follows the original masking ratio (15\%) and uses 85\% unmasked addresses to predict 15\% masked addresses, the masked addresses have a high likelihood of being present in the unmasked addresses, leading to an overly easy prediction task~\cite{he2022masked}. This issue, in turn, results in small values for loss and gradients, causing the parameters to be inadequately trained and impeding the model in capturing the meaningful occurrence patterns between addresses.

Despite the proposal of the de-duplication strategy to reduce continuous repetitiveness, the left discontinuous repetitiveness still remains substantial. To mitigate the adverse effects of high repetitiveness, we put forth two effective strategies without introducing any additional operations:

\vspace{0.1cm}
\textbf{High Masking Ratio (RR\#2):} A straightforward solution is to increase the masking ratio $k$ to a very \textit{high} value, thereby creating a task that cannot be easily extrapolated. Figure~\ref{fig:mask_ratio} demonstrates the testing $F_1$ of BERT4ETH on the phishing account detection task (with fixed-training strategy will be described in Section~\ref{sec:fixed_train}) by switching the masking ratio from 10\% to 90\%. Accordingly, the $F_1$ score increases from 0.1350 to 0.3245, causing a huge performance gap up to \textbf{18.95} AP. Among them, $k$=80$\%$ achieves superior performance, and when $k$>80$\%$, we observe that the performance starts to decrease because unmasked information is too limited for the task. 

\vspace{0.1cm}
\textbf{High Dropout Ratio (RR\#3):} An alternative approach is to adopt a \textit{high} dropout rate for pre-training, which shares the same idea with raising the masking ratio. Empirical results show that with a low masking ratio of 15\%, a similar performance can be reached by adopting a high dropout ratio of 80\%. However, the benefits brought by increasing dropout and masking ratio are not cumulative because they achieve the same effect. Therefore, given a masking ratio of 80\% adopted, a dropout ratio of 20\% is set as the default value based on empirical hyper-parameter tuning.

\subsubsection{\textbf{Skew Alleviation}}
\label{sec:skew}

As previously shown in Figure~\ref{fig:power_law}, the occurrence frequency of Ethereum accounts follows a power-law distribution, meaning that a small number of popular accounts are highly likely exists in the majority of transaction sequences, causing two irrelevant accounts be close to each other in the hidden space simply because they interact with the same popular accounts.

A good encoder is expected to identify rare activities out of the majority of transactions, as transactions that interact with low-frequency addresses could be important signals for fraud detection. We present two strategies to alleviate the negative impact of skewed distribution:

\vspace{0.05cm}
\textbf{Frequency-aware Negative Sampling (SA\#1):} Given that the masking ratio is high (80\%), compared to low-frequency addresses, high-frequency addresses are more likely to be masked and their address embeddings will be selected as $\boldsymbol{a}_{p}$ for Eq.~\ref{eq:loss}. Optimizing Eq.~\ref{eq:loss} encourages $\boldsymbol{h}_{m}$ that encodes unmasked transaction sequence be closer to $\boldsymbol{a}_{p}$ in the hidden space. As a result, addresses that co-occurr with high-frequency addresses become closer to them, and thus the sequence representations also becomes closer, which is undesired for fraud detection. An effective solution is to take high-frequency addresses as negative samples to counteract the impact of these addresses being trained frequently as positive samples. Specifically, we introduce two frequency-aware sampling strategies: Zipfan sampling and Frequent sampling, as follows:
\begin{itemize}[leftmargin=8pt]
    
\item Zipfan sampling:
\begin{equation}
    P_{neg}(A_{i}) = \frac{log(r(A_{i})+2)-log(r(A_{i})+1)}{log(max+1)}
\end{equation}
where $r(\cdot)$ is the rank of $A_{i}$ based on the descending frequency.

\item Frequent sampling:
\begin{equation}
    P_{neg}(A_{i}) = \frac{f(A_{i})^{b}}{\sum_{j}f(A_{j})^{b}}
\end{equation}
where $f(\cdot)$ is the frequency of account/address $A_{i}$ and $b$ is an adjustable hyper-parameter. In the experiment, we set $b$=0.5 and 1.0. If $b$=0, it degrades into the uniform sampling. 
\end{itemize}

\begin{table}[tbp]
\centering
\caption{Testing $F_1$ of phishing detection \textit{w.r.t.} different skew alleviation strategies. ${\dag}$ denotes intra-batch sharing.}
\setlength{\tabcolsep}{1.8mm}
\begin{tabular}{l|cccc}
\toprule
\textbf{P/N Ratio} & \textbf{1:20} &\textbf{1:1000$^{\dag}$} & \textbf{1:5000$^{\dag}$} & \textbf{1:10000$^{\dag}$} \\ 
\midrule
Uniform & 0.3245 & 0.3554 & 0.3804 & 0.3746\\
Freq(0.5) & 0.3313 & 0.3939 & 0.4214 & 0.4203 \\
Freq(1.0) & 0.3770 & 0.4390 & 0.4365 &  0.4371 \\
Zipfan & 0.4239 & 0.4251 &  \cellcolor{Gray} \textbf{0.5044}  & 0.5036 \\
\bottomrule 
\end{tabular}
\label{tab:skew_alleviation}
\vspace{0.2cm}
\end{table}

\vspace{0.1cm}
\textbf{Intra-batch Sharing (SA\#2):} Given that we sample high-frequency addresses as negative samples, the negative sets of different transaction sequences would be highly overlapped because they all concentrate on sampling high-frequency addresses. To reduce this waste, we force masked transactions in the same batch to share all the negative samples. Another advantage of this strategy is that given the whole number of negative samples unchanged, the negative/positive ratio largely increases from $|\mathcal{N}|:1$ to $ B \cdot |\mathcal{N}|: 1$, providing a greater degree of skew alleviation, where $|\mathcal{N}|$ is the size of negative set and $B$ is the batch size for transaction sequence.

Table~\ref{tab:skew_alleviation} presents the results of skew alleviation strategies on the phishing account detection task. It is obvious that: 1) BERT4ETH achieves better performance when the degree of frequent negative sampling increases (Zipfan> freq(1.0)> freq(0.5)> uniform), and the $F_1$ gap is up to \textbf{9.94} AP; 2) For Zipfan sampling, when the negative-to-positive ratio increases from 20 (without intra-batch sharing) to 1,000, 5,000 and 10,000 (with intra-batch sharing), $F_1$ increases to 0.5044, and the gap is up to \textbf{8.05} AP. As a result, the negative-to-positive ratio of 5,000 is adopted as the default setting.

\begin{figure}[tbp]
\centering
\vspace{-0.2cm}
\includegraphics[width=8.7cm]{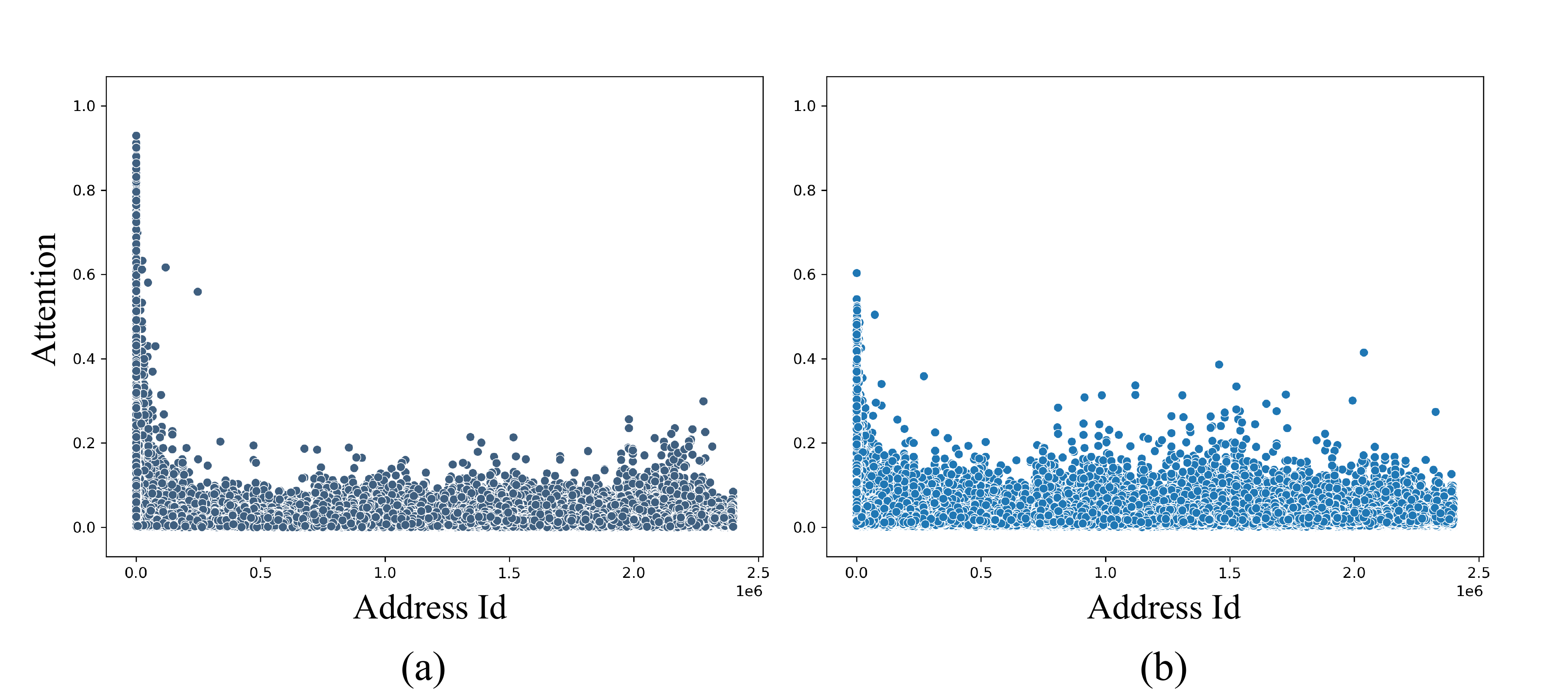}
\caption{Attention distribution with uniform(a) or Zipfan(b) negative sampling. Address Id is ranked according to the descending frequency.}
\vspace{0.05cm}
\label{fig:attention_distribution}
\end{figure}

Figure~\ref{fig:attention_distribution} demonstrate two attention distributions received by addresses in the first Transformer layer. It is obvious that attention scores assigned to the high-frequency addresses are decreased after Zipfan sampling, enabling BERT4ETH pay more attention to low-frequency addresses to enhance the distinctiveness.

\subsection{Advanced Features}
\label{sec:heterogeneity}

We designate the above-described model as the basic BERT4ETH. In this section, we present two advanced techniques aimed at tackling the heterogeneity of Ethereum transactions.

\vspace{0.05cm}
\textbf{In/Out Separation (MH\#1):} 
\label{sec:seq_separation}
It is observed that fraudulent EOAs exhibit special patterns of in-type and out-type transactions. For phishing EOAs, the in-to-out ratio is 1.250, whereas, it is only 0.385 for normal accounts. The difference arises from the nature of fraud activities: a phishing EOA receives transfers from its victims, thus making the in-type transactions dominant, whereas the out-type transactions might indicate the flow of stolen funds to accomplices or other accounts controlled by the attacker. Due to MAP is a self-supervised task, no fraud label is available to highlight these crucial yet minority transactions. As a result, these transactions may be overlooked during the self-attention computation process.

A solution is to generate other two sub-sequences by separating the original sequence based on transaction's in/out type feature. As a result, another two Transformer encoders are employed to generate $\boldsymbol{H}_{in}^{(L)}$ and $\boldsymbol{H}_{out}^{(L)}$ correspondingly. Self-attention computations are confined within each sub-sequence.
Parameters in embedding layers are shared to prevent largely increase of parameter numbers, as the address embedding layer is very large in size.

\vspace{0.1cm}
\textbf{ERC-20 Transfer Log Encoder (MH\#2):} Token transfer is implemented at the contract-level rather than at the protocol level. Therefore, a token transfer to a receiver is not recorded by an external transaction. To capture the transfer relationship, we analyze ERC-20 transfer logs from internal transactions, and select the transfer behaviors happened between EOAs to prevent noise. For an external transaction that invokes ERC-20 token transfer, we associate all the EOAs that receive ERC-20 tokens to the recipient contract address. It should be noted that the number of recipient EOAs is uncertain as a single transaction may transfer tokens to multiple EOAs simultaneously.

Before passing through the Transformer encoder, we first mean pool the address embeddings of attached EOAs to encode their information. Second, we employ a gate mechanism to integrate the embeddings of contract account and recipient EOAs:
\begin{equation}
\boldsymbol{a}_u = \operatorname{Mean}(\boldsymbol{a}_{u1}, \boldsymbol{a}_{u2}, ..., \boldsymbol{a}_{un})
\end{equation}
\begin{equation}
    \boldsymbol{a}_{c}^{'} = \boldsymbol{\beta} \cdot \boldsymbol{a}_{c} + (\mathbf{1}-\boldsymbol{\beta}) \cdot \boldsymbol{a}_u
\end{equation}
\begin{equation}
    \boldsymbol{\beta} = \operatorname{Sigmoid}(\boldsymbol{W}_{\beta}\cdot[\boldsymbol{a}_c || \boldsymbol{a}_u] + 
    \boldsymbol{b}_{\beta})
\end{equation}
where $\boldsymbol{a}_{c}$ is the address embedding of recipient contract, $\boldsymbol{\beta}$ is the gate vector that is adaptive to $\boldsymbol{a}_{c}$ and $\boldsymbol{a}_u$. $\mathbf{W}_{\beta} \in \mathbb{R}^{d \times d}$ and $ \mathbf{b}_{\beta} \in \mathbb{R}^{d}$ are parameters optimized during training. Address embedding $\boldsymbol{a}_{c}^{'}$ is encoded into the initial transaction embedding $\boldsymbol{h}_{c}^{(0)}$.
The gate mechanism prevents introducing noise in instances where transfer actions, such as token airdrops, do not necessarily guarantee the relationship between the initiator and receiver.

\subsection{Apply to Downstream Tasks}
\label{sec:apply_to_downstream}
Pre-trained BERT4ETH functions as a representation extractor for a given transaction sequence. To represent the entire sequence, we pick the representation of the self-transaction $\boldsymbol{h}_{s}^{(L)}$  because it encodes the global information. In the case where BERT4ETH adopts the in/out separation strategy, the final representation is obtained by concatenating the three representations of self-transactions extracted from $\boldsymbol{H}^{(L)}$, $\boldsymbol{H}_{in}^{(L)}$, and $\boldsymbol{H}_{out}^{(L)}$.
If the sequence exceeds the maximum length, it is split into multiple sequences and multiple representations are generated, which are then mean-pooled to produce the final representation.

\section{Experiments}

\subsection{Task Description}

\subsubsection{\textbf{Phishing Account Detection}} Phishing attack is the most prevalent form of fraud on Ethereum. Attackers send victims fake airdrop messages through emails or social networks to lure them into logging accounts on phishing websites~\cite{coindesk2022phishing} or transferring cryptocurrency to the designated phishing accounts~\cite{beetokenico}. Unlike conventional phishing scams, Ethereum transaction are publicly accessible, allowing us to identify phishing accounts and alert susceptible users before they being victimized. In our experiment, the task of phishing account detection is framed as a binary classification problem, where the goal is to determine whether an EOA is a phishing account. We adopt Precision, Recall and $F_1$ as the metrics.

\subsubsection{\textbf{De-anonymization}} De-anonymization aims to identify two different EOAs controlled by the same user. One application of de-anonymization is to trace the flow of money laundering. For example, Tornado Cash~\cite{watching,tang2021analysis} provides coin-mixing services on Ethereum: a participant deposits certain amounts of ether into a Tornado mixer contract, and use another account to withdraw the deposited coins after a period of time. In our experiment, given a ground-truth pair of EOAs, we use the representation of the query EOA to query its top-$k$ closest neighbors in the hidden space. If the target EOA is present among them, de-anonymization is considered as successful. We adopt Hit Ratio@$k$ (HR@$k$) to measure the performance and use Euclidean distance as the metric for proximity because it yield slightly better results than cosine similarity.

\begin{table}[tbp]
\centering
\caption{Dataset statistics}
\setlength{\tabcolsep}{1.1mm}
\begin{tabular}{cccccc}
\toprule
\textbf{Dataset} & \textbf{Phishing} &\textbf{ENS} & \textbf{Tornado} & \textbf{Normal} \\ 
\midrule
\textbf{\# EOA} & 3,220 & 1,335 & 2,301 & 594,038\\
\textbf{\# All address} & 151,415 & 57,526 & 139,178 & 2,609,855 \\
\textbf{\# Tranx} & 328,261 & 821,140 & 1,056,674 &  11,350,640 \\
\textbf{\# In tranx} & 182,356 & 60,258 & 150,157   & 3,159,707 \\
\textbf{\# Out tranx} & 145,905 & 760,882 & 906,517 & 8,190,933 \\
\bottomrule 
\end{tabular}
\label{tab:statistic}
\vspace{0.1cm}
\end{table}

\subsection{Experimental Setup}

\subsubsection{\textbf{Dataset:}} For phishing account detection, we collect 7,057 accounts labeled by Etherscan. Among them, 97\% are EOAs. For de-anonymization, we use ground-truth pairs collected by Xu et al.~\cite{watching} from two sources: Ethereum Name Service (ENS) and Tornado Cash.
For pre-training, we randomly sample 1,000,000 EOAs and filter out accounts labeled as "phishing", "exchange", "miner" and "mining pool". These normal EOAs are also used for negative samples for phishing account detection and candidate set for de-anonymization on ENS dataset.
We collect all the transactions that these EOAs has involved in, covering the period from Jan.1 2017 to May.1 2022. We filter out EOAs that has less than 3 transactions or more than 10,000 transactions. The statistics of dataset is presented in Table~\ref{tab:statistic}.

\subsubsection{\textbf{Baselines:}} We compare three types of baselines in the experiment: 1) DeepWalk-based methods, including DeepWalk~\cite{deepwalk}, Trans2Vec~\cite{trans2vec}, Diff2Vec~\cite{diff2vec} and Role2Vec~\cite{role2vec}. Trans2Vec is specifically designed for phishing account detection. Diff2Vec is the SOTA method for de-anonymization in~\cite{watching}; 2) GNN-based methods, including GCN~\cite{gcn}, GraphSAGE~\cite{graphsage} and GAT~\cite{gat}; 3) BERT4ETH and its variants, including the basic BERT4ETH, BERT4ETH$^{\dag}$ with in/out separation and BERT4ETH$^{\S}$ with ERC-20 log encoder.

\subsubsection{\textbf{Implementation:}} 

For BERT4ETH, the number of Transformer layers is set to 8, number of attention head is set to 2 and maximum sequence length is set to 100.
For graph-based methods, we adopt the self-supervised training task proposed in DeepWalk~\cite{deepwalk}. We set the number of walks per node to 10, walk length to 20, and context size to 5. For GNN-based methods, the number of GNN layers is set to 2 with the neighbor sample size of 50. For all the competitors, the negative-to-positive ratio is set to 20, hidden dimension set to 64, batch size set to 256 and dropout ratio set to 20\%. Other parameters of BERT4ETH follows the previously mentioned default settings.

\subsection{Performance Comparison}
\label{sec:fixed_train}
\subsubsection{\textbf{Phishing Account Detection:}} We evaluate BERT4ETH \textit{w.r.t.} two strategies: fixed-training and fine-tuning. For fixed-training, the pre-trained model is used a feature extractor to generate representations, followed by the individual training of a MLP for classification. For fine-tuning, the model is trained with a cascaded MLP together. Each experiment is repeated five times and the best $F_1$ score is reported with the threshold set to 0.3.

\begin{figure}[tbp]
\centering
\includegraphics[width=8.6cm]{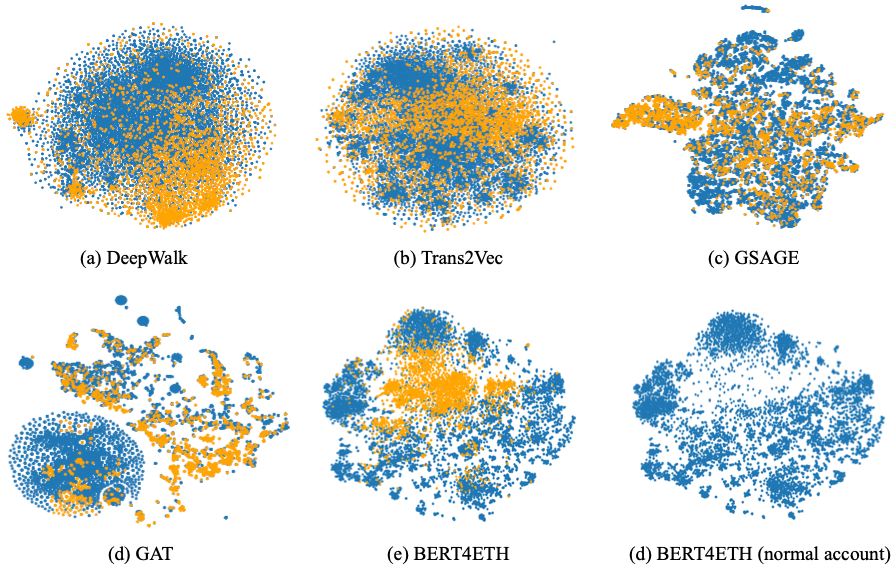}
\caption{T-SNE visualization of phishing (orange) and normal (blue) accounts for several competitors. }
\vspace{0.05cm}
\label{fig:tsne_visualization}
\end{figure}

Table~\ref{tab:phishing_fixed_train} summarizes the results of the fixed-training strategy. From the table we can observe that: 1) GNN-based methods, especially GAT, outperform DeepWalk-based methods.
2) The basic BERT4ETH achieves a significant performance boost compared to other baselines: the performance gap of $F_1$ is up to \textbf{21.61} AP compared to GNNs, and \textbf{36.94} AP compared to the original BERT with $F_1$ of 13.50 (Figure~\ref{fig:mask_ratio}). The performance boost mainly comes from a better pre-training process, incorporating fine-grained sequential and transaction information, as well as the superior modeling ability of Transformer. 3) By applying two advanced features, BERT4ETH$^{\dag}$ and BERT4ETH$^{\S}$ further introduce \textbf{3.44} and \textbf{1.68} AP gain of $F_1$, respectively. 
The comparison of account representations generated by several baselines, as visualized in Figure~\ref{fig:tsne_visualization}, show that phishing account representations generated by BERT4ETH are more dense and separable.

Table~\ref{tab:phishing_finetune} presents the results after fine-tuning, with graph-based methods omitted as they perform worse and some of them cannot be fine-tuned. The first three rows of Table~\ref{tab:phishing_finetune} show the results of fine-tuning the pre-trained BERT4ETH models, showing a substantial improvement in performance compared to Table~\ref{tab:phishing_fixed_train}. This indicates the effectiveness of fine-tuning. The last three rows are results of ablating the pre-training. Obviously, the performance largely decreases, suggesting the importance of pre-training. Among these results, BERT4ETH$^{\dag}$ achieves the best performance.

\begin{table}[tbp]
\centering
\caption{\small{Comparison for phishing detection w/ fixed-training.}}
\setlength{\tabcolsep}{2.8mm}
\begin{tabular}{l|ccc}
\toprule
\textbf{Method}    & \textbf{Precision} &  \textbf{Recall}  & \textbf{F$_1$}     \\ \midrule
DeepWalk  & 0.2293 & 0.1734  &  0.1974  \\
Trans2Vec & 0.1636  & 0.1432  &  0.1527  \\ 
Diff2Vec  & 0.2184 & 0.2000  &  0.2088  \\
Role2Vec  & 0.2520  & 0.2295  &  0.2402  \\ 
GCN       &  0.3181  &  0.2180  &  0.2587 \\
GSAGE     &  0.3023 &  0.2317  &  0.2623 \\
GAT       &  0.3264  &  0.2581  & 0.2883 \\
\midrule 

\rowcolor{Gray}
BERT4ETH  & 0.5483  &  0.4670  &  0.5044 \\

BERT4ETH$^{\dag}$ & 0.5826  &  \textbf{0.5012}  & \textbf{0.5388}  \\
BERT4ETH$^\S$ & \textbf{0.5858} & 0.4695 & 0.5212  \\

\bottomrule
\end{tabular}
\label{tab:phishing_fixed_train}
\vspace{0.2cm}
\end{table}

\begin{table}[tbp]
\centering
\caption{\small{Comparison for phishing detection w/ fine-tuning. BERT4ETH w/o pre-training equals to Transformer.}}
\setlength{\tabcolsep}{2.8mm}
\begin{tabular}{l|cccc}
\toprule
\textbf{Method}    & \textbf{Precision} &  \textbf{Recall}  & \textbf{F$_1$}    \\ \midrule
\rowcolor{Gray}
BERT4ETH  & 0.7153 & 0.5984 & 0.6516 \\
BERT4ETH$^{\dag}$ & \textbf{0.7421} & \textbf{0.6125} & \textbf{0.6711} \\
BERT4ETH$^\S$ & 0.7162 &  0.6107  &  0.6593 \\

\midrule
\multicolumn{4}{c}{\textbf{w/o pre-training}}  \\ 
\midrule
\rowcolor{Gray}

BERT4ETH  &  0.5114  &  0.3845  &  0.4389  \\
BERT4ETH$^{\dag}$ & \textbf{0.5287}  &  \textbf{0.4129}  & \textbf{0.4637}  \\
BERT4ETH$^\S$ & 0.5110 & 0.4058 & 0.4524  \\

\bottomrule
\end{tabular}
\vspace{0.2cm}
\label{tab:phishing_finetune}
\end{table}

\subsubsection{\textbf{De-anonymization:}} For ENS dataset, we construct a candidate set including ENS and normal EOAs (595,373 in total), which is shared by all the ground-truth pairs. For Tornado dataset, we use ground-truth pairs collected from 0.1 ETH and 1 ETH coin-mixers. Each ground-truth pair consisting of a deposit EOA and a withdraw EOA, we construct a candidate set including EOAs that deposited Ether to the mixers prior to the withdrawal time. The withdraw EOA is used to query the deposit EOA within the candidate set. BERT4ETH cannot be fine-tuned for this task due to the limited number of ground-truth pairs, with 288 pairs for the ENS dataset and 182 pairs for the Tornado dataset.

Table~\ref{tab:ens_result} presents the comparison results for the ENS dataset. Among the DeepWalk-based methods, Diff2Vec demonstrates the best performance. Among GNNs, GraphSAGE outperforms both GCN and GAT by a large margin, which can be attributed to the fact that homogeneous GNNs can introduce a large amount of noise in the multi-hop aggregation, which de-anonymization is susceptible to, especially given that Ethereum account nodes that are highly likely to be linked to popular account nodes. GraphSAGE is more resistant to noise~\cite{Turbo} because of its skip-connection mechanism. Notably, the basic BERT4ETH can exactly de-anonymize 16.32\% of account pairs, offering a significant improvement of \textbf{9.7} AP gain on HR@1. Additionally, the in/out separation strategy further brings a considerable improvement upon the basic BERT4ETH.

Table~\ref{tab:tornado_result} presents the results for the Tornado dataset, where Rank is the average rank of the deposit account within the candidate set. The size of candidate set varies due to the unique withdrawal time. Similarly, BERT4ETH significantly outperforms its competitors. The results show that it can even exact de-anonymize \textbf{40.2\%} of testing pairs for 0.1 ETH mixer and \textbf{45.0\%} of for 1 ETH mixer.

\begin{table}[tbp]
\centering
\caption{\small{Comparison for de-anonymization on the ENS dataset (\%).}}
\setlength{\tabcolsep}{1.05mm}
\begin{tabular}{l|ccccccc}
\toprule
\textbf{Method} & \textbf{HR@1}  & \textbf{@3} &\textbf{@5} &\textbf{@10} &\textbf{@100} &\textbf{@1000} \\ 
\midrule
DeepWalk &  2.43 & 5.21 & 5.56 & 7.64 & 15.28 & 41.32 \\
Trans2Vec & 6.60 & 8.33 & 9.03 & 11.46 & 19.44 & 24.65  \\
Diff2Vec  & 3.82 & 5.90 & 6.25  & 9.03 & 22.22 & 43.75  \\
Role2Vec & 4.17 & 5.56 & 6.25 & 7.30 & 17.71 & 35.07 \\
GCN     & 1.74 & 2.78  & 3.13 & 4.51  & 11.46 & 31.94 \\
GSAGE   & 5.90 & 9.03  & 10.07 & 12.85 & 26.39 & 55.90 \\
GAT     & 3.13 & 4.51  & 5.21 & 6.25  & 18.40 & 47.57 \\
\midrule
\rowcolor{Gray}
BERT4ETH & 16.32 & 23.26 & 28.47 & 32.64 & 47.92 & 65.28 \\
BERT4ETH$^{\dag}$ & \textbf{20.14} & \textbf{27.43} & \textbf{31.60} & \textbf{36.11} & \textbf{51.39} & \textbf{68.75} \\
BERT4ETH$^{\S}$ & 18.40 & 25.69 & 29.86 & 34.03 & 50.00 & 67.71 \\

\bottomrule         
\end{tabular}
\label{tab:ens_result}
\end{table}

\begin{table}[tbp]
\small
\centering
\caption{\small{Comparison for de-anonymization on the Tornado dataset.}}
\setlength{\tabcolsep}{1.1mm}
\begin{tabular}{l|ccc|ccc}
\toprule
\textbf{Mixer}     & \multicolumn{3}{c|}{\textbf{0.1 ETH}}                   & \multicolumn{3}{c}{\textbf{1 ETH}}                     \\ \midrule
          & \multicolumn{2}{c}{ \textbf{\# Candidate}} & \textbf{\# Pair}   & \multicolumn{2}{c}{\textbf{\# Candidate}} & \textbf{\# Pair} \\ \midrule
Statistic & \multicolumn{2}{c}{405.2}              &  102   &  \multicolumn{2}{c}{263.7}              &     80     \\ \midrule
\textbf{Method}    & \textbf{Rank$\downarrow$}            & \textbf{HR@1}            & \textbf{HR@3}      & \textbf{Rank$\downarrow$}            & \textbf{HR@1}            & \textbf{HR@3}     \\ \midrule
DeepWalk  &      112.64    &    17.65\%     &   20.59\%   &   63.59 &   26.25\%   &    37.50\%     \\
Trans2Vec &  100.69    &    18.63\% &   28.43\%  & 70.57  &  26.25\% &  35.00\%   \\
Diff2Vec  &   82.99  & 26.47\%  & 37.25\% & 54.00  &  35.00\%  & 43.75\% \\
Role2Vec &  100.99 & 21.57\% &  30.39\%  &  64.65   &   15.00\%  & 31.25\% \\ 
GCN       &   153.75    &   13.73\%   &   17.65\%     &   87.01   &   16.25\%    &   25.00\%      \\
GSAGE     &    89.46       &   30.39\%       &    37.25\%   &    58.01     &    35.00\%       &    48.75\%      \\
GAT       &    91.25     &    21.57\%      &   27.45\%       &   60.42    &  25.00\%               &    38.75\%    \\ \midrule 

\rowcolor{Gray}
BERT4ETH & 67.96 & 40.20\% & 56.86\% & 42.26 & 45.00\% & 62.50\% \\
BERT4ETH$^{\dag}$ & \textbf{62.56} & \textbf{51.96}\% & \textbf{58.64}\% &  \textbf{37.34} & \textbf{53.75}\% & \textbf{66.25}\% \\
BERT4ETH$^{\S}$ & 67.17 & 39.22\% & 55.88\% &  41.66 & 46.25\% & 62.50\% \\

\bottomrule

\end{tabular}
\label{tab:tornado_result}
\vspace{0.2cm}
\end{table}



\subsection{Ablation Study} 

We investigate the impact of all the proposed strategies by ablating five key elements of BERT4ETH, \textit{i.e.}, transaction de-duplication, high masking ratio, frequent negative sampling, intra-batch sharing and transaction features.

Table~\ref{tab:ablation_phishing} presents the results of the ablation study conducted on the phishing account detection task with fixed training. First, we observe that removing each one of them results in a noticeable performance decline. Second, we observe that switching masking ratio to BERT's original setting (15\%) lead to the largest performance decrease, indicating that repetitiveness can largely hurt the effectiveness of pre-training for the phishing detection task.

On the contrary, the conclusion drawn from the ablation study on the de-anonymization task is entirely different. Ablating repetitiveness reduction strategies (de-duplication \& 80\% masking) actually lead to a slight performance increase when $k\leq5$, suggesting that de-anonymization task is not particularly susceptible to high repetitiveness. Another noteworthy finding is that two skew alleviation strategies are crucial for this task. It is worth mention that skew alleviation is important especially when the masking ratio is high. This is because high-frequency addresses are more likely be selected as the positive addresses, which can cause the whole sequence representation to be closer to them, making the representations indistinguishable.

\begin{table}[tbp]
\centering
\caption{\small{Ablation study for phishing detection w/ fixed-training.}}
\setlength{\tabcolsep}{2.8mm}
\begin{tabular}{l|cccc}
\toprule
\textbf{Method}    & \textbf{Precision} &  \textbf{Recall}  & \textbf{F$_1$}    \\ \midrule
BERT4ETH  & 0.5483 &  0.4670 & 0.5044  \\
w/o de-duplication & 0.4661 &  0.3289  & 0.3857  \\
w/o 80\% masking & 0.4177   & 0.2431   & 0.3073  \\
w/o freq. sampling &  0.4554  &  0.3266  &  0.3804 \\
w/o batch-sharing &  0.5034  &  0.3661  &  0.4239 \\
w/o tranx. features &  0.5076  &  0.3607  &  0.4217 \\

\bottomrule
\end{tabular}
\label{tab:ablation_phishing}
\end{table}

\begin{table}[tbp]
\centering
\caption{\small{Ablation study for de-anonymization on ENS dataset (\%).}}
\setlength{\tabcolsep}{0.8mm}
\begin{tabular}{l|cccccc}
\toprule
\textbf{Method}    & \textbf{HR@1} &  \textbf{@3}  & \textbf{@5} & \textbf{@10} & \textbf{@100} & \textbf{@1000} \\ \midrule
BERT4ETH & 16.32 & 23.26 & 28.47 & 32.64 & 47.92 & 65.28 \\
w/o de-duplication & 18.06 & 26.39 & 30.21 & 32.29 & 46.18 & 59.38 \\
w/o 80\% masking  & 16.67 & 23.96 & 27.43 & 31.60 & 41.33 & 52.78 \\
w/o freq. sampling & 4.52 & 9.03 & 10.07 & 11.11 & 27.78 & 48.61\\
w/o batch-sharing  & 0.69 & 1.74 & 3.82 & 5.90 & 21.88 & 43.40\\
w/o tranx. features & 15.28 & 21.53 & 26.74 & 27.09 & 42.71 & 62.85 \\

\bottomrule
\end{tabular}
\label{tab:ablation_ens}
\end{table}

\begin{figure}[htbp]
\centering
\vspace{0.3cm}
\includegraphics[width=7cm]{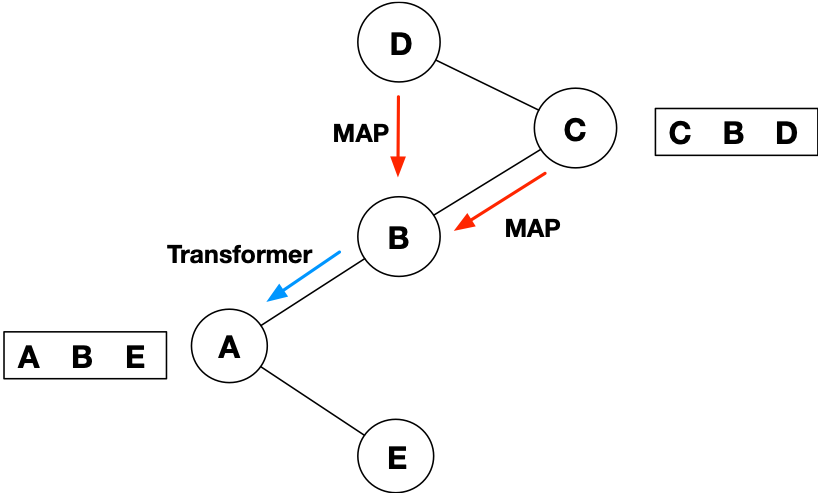}
\vspace{0.2cm}
\caption{A toy example illustrating that BERT4ETH can capture three-hop neighborhood information.}
\vspace{-0.1cm}
\label{fig:multihop_modeling}
\end{figure} 

\subsection{Multi-hop Modeling}
\label{sec:multihop_modeling}

Despite modeling at transaction sequence level, BERT4ETH can still capture up to three-hop neighborhood information from a graph perspective. As illustrated in Figure~\ref{fig:multihop_modeling}, we represent five accounts as nodes and transactions between them as edges. When conducting MAP pre-training on the sequence of node C, the address embedding of node B becomes closer to the address embeddings of nodes C and D, suggesting that address embeddings trained during MAP implicitly incorporate information from up to two-hop neighborhood. After pre-training, we use a Transformer encoder to extract the account representation of node A by taking its transaction sequence as input, which explicitly incorporates the address embedding of B and, implicitly, the information from nodes C and D. By combining MAP and the Transformer, BERT4ETH is able to capture a total of \textbf{three-hop} neighborhood information.

\begin{table}[htbp]
\centering
\caption{\small{Comparison for phishing detection w/ fixed-training.}}
\setlength{\tabcolsep}{2.8mm}
\begin{tabular}{l|ccc}
\toprule
\textbf{Method}    & \textbf{Precision} &  \textbf{Recall}  & \textbf{F$_1$}     \\ \midrule
BERT4ETH  & 0.5483  &  0.4670  &  0.5044 \\
\rowcolor{Gray}
Addr Embed  & 0.3775  &  0.2604  &  0.3084 \\
\bottomrule
\end{tabular}
\label{tab:multihop_modeling_phishing}
\end{table}

\begin{table}[htbp]
\centering
\caption{\small{Comparison for de-anonymization on the ENS dataset (\%).}}
\setlength{\tabcolsep}{1.05mm}
\begin{tabular}{l|ccccccc}
\toprule
\textbf{Method} & \textbf{HR@1}  & \textbf{@3} &\textbf{@5} &\textbf{@10} &\textbf{@100} &\textbf{@1000} \\ 
\midrule
BERT4ETH & 16.32 & 23.26 & 28.47 & 32.64 & 47.92 & 65.28 \\
\rowcolor{Gray}
Addr Embed & 13.32 & 19.26 & 24.47 & 25.64 & 30.92 & 39.28 \\
\bottomrule         
\end{tabular}
\label{tab:multihop_modeling_ens}
\end{table}

After pre-training, we extract the address embeddings to directly represent accounts (addresses) and then test them on two tasks. Table~\ref{tab:multihop_modeling_phishing} and Table~\ref{tab:multihop_modeling_ens} present the corresponding results, from which we notice that while the performance declines compared to the BERT4ETH representations, address embeddings remain effective for both tasks. This indicates that the address embeddings pre-trained during MAP are already equipped to capture multi-hop information. Furthermore, we observe a more significant performance drop in the phishing account detection compared to the de-anonymization task, implying that multi-hop information is more crucial for the former task.

\begin{table}[htbp]
\small
\centering
\vspace{0.2cm}
\caption{\small{Case study of hot-to-cold query for de-anonymization.}}
\setlength{\tabcolsep}{1.1mm}
\begin{tabular}{l|ccc|ccc}
\toprule  

 & \multicolumn{3}{c|}{\texttt{Case 1}} & \multicolumn{3}{c}{\texttt{Case 2}} \\ 
\midrule
\textbf{Query type} & \multicolumn{2}{c}{ \textbf{cold-to-hot}} & \textbf{hot-to-cold}   & \multicolumn{2}{c}{\textbf{cold-to-hot}} & \textbf{hot-to-cold} \\ \midrule
Diff2Vec & \multicolumn{2}{c}{8th}              &  47,641st   &  \multicolumn{2}{c}{10th}              &     49,059th     \\ 
BERT4ETH & \multicolumn{2}{c}{8th}              &  20th   &  \multicolumn{2}{c}{2th}              &     54th     \\ 
\bottomrule
\end{tabular}
\label{tab:tornado_result}
\end{table}

\subsection{Case Study}

We identify the case of hot-to-cold query for de-anonymization where graph-based methods may fail. In this case, the cold account has much less number of transactions than the hot account. Take \texttt{Case 1} for example, when using the cold account to query the hot account, both BERT4ETH and Diff2Vec rank the hot account at 8th/595,396. However, when using the hot account to query the cold account, BERT4ETH ranks 20, and Diff2Vec ranks 47,641. The reason is that graph-based methods can introduce a large amount of noise when a node’s neighborhood is large. In contrast, BERT4ETH preserves the first order of neighborhood with transaction-level sequential information, and emphasizes important information via self-attention, making it still effective for hot-to-cold query.

\vspace{0.1cm}
 \noindent \texttt{Case 1}:

\noindent \textbf{Cold EOA}: 0x92fa836a964fd5544545b18285475f715acb5576 (7)

\noindent \textbf{Hot EOA}: 0xf199b022e3edab7e8ac6e214fad1bdfd31703766 (922)

\vspace{0.1cm}
 \noindent \texttt{Case 2}:
 
\noindent \textbf{Cold EOA}: 0x53c5438e8c825ba574865b4c1ccb34e74b3affdf (5)

\noindent \textbf{Hot EOA}: 0xd9ceb2bb4b8324d36c284799eb00c7cc19a0f618 (143)


\section{Conclusion}

We present BERT4ETH, a pre-trained Transformer that offers a universal solution for fraud detection tasks on Ethereum. BERT4ETH features the superb modeling ability of the Transformer and incorporates three effective strategies to tackle the challenges of pre-training a BERT model for Ethereum. These strategies, namely repetitiveness reduction, skew alleviation and heterogeneity modeling, result in substantial improvements and operate cohesively and harmoniously. The significant improvements achieved on phishing account detection and de-anonymization tasks suggest that BERT4ETH is well suited for practical applications.

\section{Acknowledgement}

This research was supported by the National Research Foundation, Singapore under its Industry Alignment Fund – Pre-positioning (IAF-PP) Funding Initiative. Any opinions, findings and conclusions or recommendations expressed in this material are those of the author(s) and do not reflect the views of National Research Foundation, Singapore.
Authors of Georgia Tech acknowledge the partial support by the US National Science Foundation under Grants NSF 1564097, NSF 2026945, NSF 2038029, a CISCO grant and an IBM Faculty Award.


\bibliographystyle{abbrv}
\bibliography{BERT4ETH}


\end{document}